\documentclass[12pt]{iopart}
\usepackage{epsf}
\usepackage{epsfig}
\usepackage{graphicx}% Include figure files
\usepackage{dcolumn}% Align table columns on decimal point
\usepackage{bm}% bold math
\usepackage{iopams}  
\usepackage{hyperref}

\begin{document}

\title[Performance enhancement of TiO$_2$-based DSSCs by carbon nanospheres]{Performance enhancement of TiO$_2$-based dye-sensitized
solar cells by carbon nanospheres in photoanode}.
%\title{Performance enhancement of TiO$_2$-based dye-sensitized solar cells by carbon nanospheres in photoanode}

\author{Elham Bayatloo, and Esmaiel Saievar-Iranizad}
\address{Department of Physics, Tarbiat Modares University, P.O. Box 14115-175 Tehran, Iran}
\eads{\mailto{e.bayatloo@yahoo.com}, \mailto{saievare@modares.ac.ir}}
%\ead{saievare@modares.ac.ir}

\date{\today}

\begin{abstract}
The conversion efficiency of dye-sensitized solar cells (DSSCs) is optimized by modifying the optical design
and improving absorbance within the cell. 
These objectives are obtained by creating different sized cavities in TiO$_2$ photoanode. For this purpose,
carbon nanospheres with diameters $100$-$600$ nm are synthesized by hydrothermal method. A paste of 
TiO$_2$ is mixed with various amounts of carbon nanospheres. 
During TiO$_2$ photoanode sintering processes at $500^{\circ}$C temperature, 
the carbon nanospheres are removed.
This leads to random creation of cavities in the DSSCs
photoanode. These cavities enhance light scattering and porosity which improve 
light absorbance by dye $N719$ and provide a larger surface area for dye loading.
These consequences enhance performance of DSSCs. By mixing $3\%$ Wt.
carbon nanospheres in the TiO$_2$ pastes, we were able to increase the short
circuit current density and efficiency by $40\%$ (from $12.59$ to $17.73$ mA/cm$^2$)
and $33\%$ (from $5.72\%$ to $7.59\%$), respectively.

\end{abstract}
\submitto{\NT}
\maketitle

\section{introduction}
Since the fabrication of dye-sensitized solar cells (DSSCs) in $1991$~\cite{Gratzel91}, many researches have
been done for enhancing the performance and stability of these cells. During the last two decades, DSSCs have taken a lot of
attention because of their low cost and ease of manufacturing. 
The efficiency of DSSCs can be influenced by various characters
such as charge recombination~\cite{reducerecombination1,reducerecombination2}, light absorption by dye~\cite{absorbance},
specific surface area and porosity~\cite{porosity}, electron transport~\cite{chargetransport}
and so on.
These parameters can be controlled by a variety of
approaches like using different dyes~\cite{newdye}, semiconductors~\cite{semi} or electrolytes~\cite{electrolyte}, 
changing morphology~\cite{morpho} or band gap of semiconductors~\cite{bandgap}, photon management for light
scattering~\cite{Usami1997,Usami2000,smallandlarge}, etc~\cite{dyeinelectrolyte,mixdye}. 
Among these methods, photon management is used to enhance the photon path length in order to increase
the probability of light absorption by dye. 

The most successful implementation of photon management in DSSCs is usage of diffuse scattering layer~\cite{scatterer}.
This scattering layer, which is deposited on the top of photoanode main layer,
is consisted of large size particles with sizes between $300$-$1000$ nm. These particles
are made of transition metal oxide with high refractive index such as TiO$_2$. 
In recent times, disordered structures such as one, two or three dimensional photonic crystals ($1$DPC,
$2$DPC, $3$DPC)~\cite{PC1,PC2} and TiO$_2$ hollow spheres~\cite{Dadgostar12} have been used as scattering layer
instead of TiO$_2$ filled spheres.
The replacement of TiO$_2$ hollow spheres with TiO$_2$ filled spheres has leaded to increment of light absorbance
as is shown in Ref.~[20].

In another implementation of photon management, a mixture of both small and large particles is employed in photoanode
in order to scatter light~\cite{mix}.
Usage of porous structures in photoanode is another possibility for light scattering and improving dye absorption~\cite{porosity2}.
Many researches have been carried out to create porosities and cavities in the photoanode thin film. Recently,
polystyrene ball embedded in a paste of TiO$_2$ has created cavities during sintering process. The role of these
cavities is to scatter light and increase light absorption by dye~\cite{polystyrene}.

In this research, carbon nanospheres powder was synthesized by hydrothermal method that is a facile
and low cost method. Different amounts of carbon nanospheres were mixed with a paste of TiO$_2$ nanostructure
consisted of $\sim$ $20$ nm TiO$_2$ nanocrystals as describe below. Consequentially, effect of TiO$_2$ porosity in photoanode 
on absorption of dye, light scattering, charge recombination, and chemical capacitance
are investigated. Ultimately, an energy conversion efficiency up to
$7.59\%$ achieved for $3$ $\%$Wt. of carbon nanospheres mixed with TiO$_2$.  

\maketitle

\section{Experimental details}
\subsection{Synthesis of carbon nanospheres}
Synthesis of carbon nanospheres was done with polycondensation reaction of glucose under
hydrothermal conditions described in Ref.~\cite{carbon}. A $0.5$ M aqueous solution of glucose was prepared. $50$
mL of this solution was kept in $60$ mL Teflon lined autoclave and heated for $14$ h at $160^{\circ}$C. The black
or brown products were centrifuged and washed with ethanol and water three times and dried at 80$^{\circ}$C
for more than $4$ h. 
\subsection{Mixing TiO$_2$ paste and carbon nanospheres powder}
For mixing carbon nanospheres powder and commercial TiO$_2$ paste consisting of $\sim$ $20$ nm TiO$_2$
nanoparticles, carbon nanospheres were dissolved in ethanol to obtain $1$, $2$, $3$, and $4$ Wt.$\%$ solutions.
These solutions were sonicated to disperse completely and then mixed with TiO$_2$ paste with a weight
ratio of $1$:$2$. The resultant mixture was dispersed with ultrasonic titanium probe and then was
concentrated with rotary-evaporator to remove ethanol. Five different TiO$_2$ pastes were prepared with
various weight percentages of ethanolic solutions of carbon nanospheres about $0\%$, $1\%$, $2\%$, $3\%$, and
$4\%$ and named T$0$, T$1$, T$2$, T$3$ and T$4$, respectively. 
\subsection{Fabrication of TiO$_2$ photoelectrode}
For preparing DSSC working electrode, FTO glass ($3$ mm thickness, $15\Omega$/$\Box$, Dyesol) was first washed
with detergent and rinsed with DI water. Then it was cleaned by using an ultrasonic bath with DI water, $0.1$ M
HCl solution in ethanol, acetone, and ethanol for $15$ min each. The washed FTO was heated up to $450^{\circ}$C for $15$ min.
Afterwards a compact blocking layer of TiO$_2$ was deposited onto FTO by immersing FTO into $40$ mM aqueous TiCl$_4$
solution for $30$ min at $70^{\circ}$C and washed with DI water and ethanol. The next layer was a porous TiO$_2$ layer
(T$0$, T$1$, T$2$, T$3$ or T$4$) deposited by doctor blade technique with $\sim$ $10$ $\mu$m thickness. After drying the
doctor bladed films for $6$ min at $125^{\circ}$C, the films were heated at $325^{\circ}$C for $5$ min, at $375^{\circ}$C for
$5$ min, at $450^{\circ}$C for $15$ min, and at last at $500^{\circ}$C for $30$ min. Finally, for depositing last layer, the films
again were treated in $40$ mM aqueous TiCl$_4$ solution as described previously and sintered at $500^{\circ}$C for $30$ min.
After cooling naturally to $80^{\circ}$C temperature, TiO$_2$ electrodes were immersed into a $0.4$ mM N$719$ (Dyesol)
dye solution in ethanol for $20$-$24$ h.
In this research five different working electrodes (T$0$, T$1$, T$2$, T$3$, and T$4$) were prepared corresponding to
the different TiO$_2$ pastes.

\subsection{Preparation of counter electrode}
For preparing DSSC counter electrode, a hole with diameter about $6$-$7$ mm was drilled on the FTO glass. Then the FTO glass
was washed as described above for FTO working electrode but instead for $6$ min. After that, the FTO was heated to $470^{\circ}$C
for $15$ min. The thermal decomposition method was employed to deposit the Pt catalyst on the FTO glass. Whit this purpose, a
drop of $0.3$ mM H$_2$PtCl$_6$ solution in ethanol was applied on the FTO glass and annealed at $470^{\circ}$C for $15$ min.

\subsection{Electrolyte}
The electrolyte I$^-$/${\rm I}^{-}_3$ consisted of $0.1$ M LiI, $0.1$ M I$_2$, $0.5$ M $4$-tert-butylpyridine, and $0.6$ M
Tetrabutylammonium iodide in acetonitrile.

\subsection{DSSC assembly}
The dye loaded porous TiO$_2$ photoanode and the Pt counter electrode were sealed together with a $30$ $\mu$m
Surlyn (Dyesol) spacer around the TiO$_2$ active area ($0.25$ cm$^2$). The liquid electrolyte (I$^-$/${\rm I}^{-}_3$) was injected
into assembled cells. At the end, the hole of counter electrode was covered by a glass ($1$cm*$1$cm) and sealed by a spacer.

The morphology of carbon nanospheres and TiO$_2$ nanoparticles and the thickness of the layer were observed by field-
emission scanning electron microscope (Hitachi S-$4160$). The current-voltage (I-V) characteristics of the
fabricated DSSCs were measured under solar simulator illumination of AM $1.5$ ($100$ mW/cm$^2$). The
electrochemical impedance spectroscopy (EIS) measurements of the cells were performed with an
Iviumstat. The EIS measurements were carried out in dark conditions and in room temperature, by applying
a sinusoidal potential perturbation of 10 mV with the frequency ranging between $1$ MHz and $0.05$ Hz at
different applied forward biases. Incident photon to current conversion efficiency (IPCE) spectra were
recorded on a Jarrell Ash monochromator using a $100$ W halogen lamp and a calibrated photodiode
(Thorlabs). Diffuse reflection spectra of the films have been collected by Avaspec$2048$-TEC UV-Vis-NIR spectrophotometer with integrating sphere Avaligth-DHS. UV-Vis spectra of the dye loaded TiO$_2$
films were determined with a UV-Vis spectrophotometer (PG instrument, T$80$+). The concentration of
absorbed dyes was determined by first desorbing the dyes from the dye-sensitized TiO$_2$ films in $0.1$ M NaOH
aqueous solution and then analyzing by UV-Vis spectrophotometer.

\begin{figure}
  \begin{center}
    \includegraphics[angle=0,width=8.5cm]{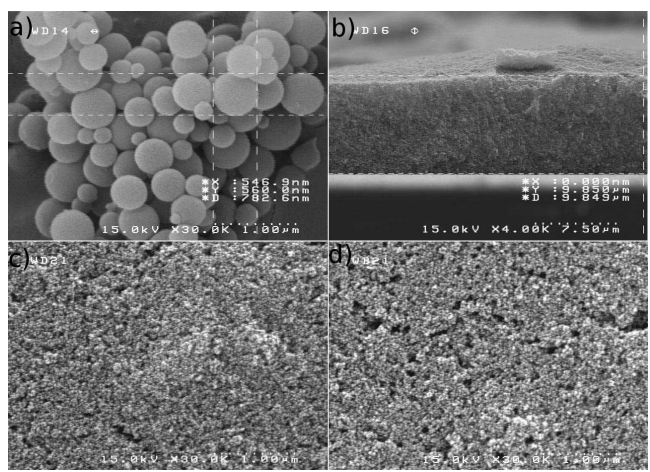}
    \caption{FE-SEM images of a) carbon nanospheres b) cross section of TiO$_2$ thin film c) top view of TiO$_2$ photoanode without
    carbon nanospheres (T$0$) and d) with $3$ $\%$ Wt. carbon nanospheres (T$3$)}
    \label{sem}
  \end{center}
\end{figure}

\section{RESULTS AND DISCUSSION}
\begin{figure}[b]
  \begin{center}
    \includegraphics[angle=-90,width=9cm]{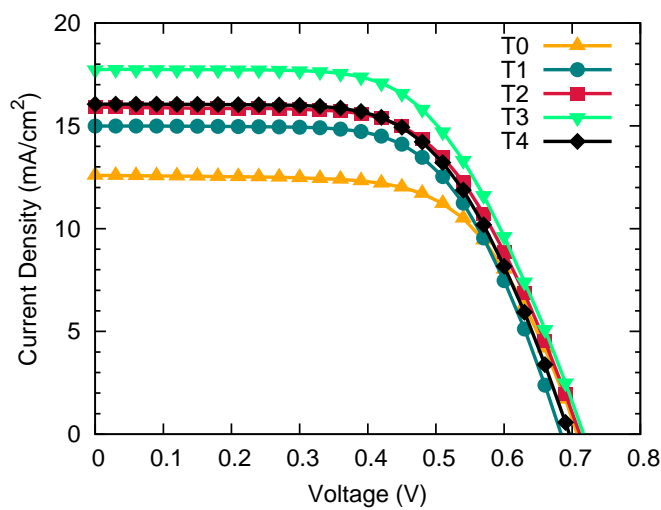}
    \caption{I-V curves of DSSCs fabricated with TiO$_2$ and TiO$_2$ mixed carbon nanospheres
    with different weight percentages $1$, $2$, $3$ and $4$ $\%$Wt (T$0$, T$1$, T$2$, T$3$, and T$4$ cells).}
    \label{Efficiency}
  \end{center}
\end{figure}

\fref{sem} shows field-emission scanning electron microscopy (FE-SEM) images of carbon nanospheres (a), as well as cross section
(b) and top view (c,d) of TiO$_2$ thin film. In figure~1a, the spherical morphology with the size range of $100$-$600$ nm is indicated.
figure~1b shows cross sectional FE-SEM image of the TiO$_2$ thin film deposited on FTO substrate. This Figure confirms that
the TiO$_2$ thin film had the thickness of about $10$~$\mu$m.
figure~1c and d show the top view FE-SEM images of the sintered TiO$_2$ with and without carbon nanospheres for
T$0$ and T$3$ photoanodes. 
These images make the point clear that adding carbon nanospheres to TiO$_2$ paste increases porosity of TiO$_2$ thin films.
During TiO$_2$ sintering processes, carbon nanospheres were removed at $490^{\circ}$C and leaded to
the creation of cavities ~\cite{Dadgostar12}. These cavities enlarged the surface area and thus increased the
number of absorbed dye molecules (N$719$ dye) on the TiO$_2$ thin film.
The I-V curves of the different DSSCs fabricated with the T$0$, T$1$, T$2$, T$3$, and T$4$ photoanodes are depicted
in \fref{Efficiency}. Table~\ref{table} lists photovoltaic parameters of the T$0$, T$1$, T$2$, T$3$, and T$4$ DSSCs.
As seen in Table~\ref{table}, the absorbed dye molecules for T$3$ photoanode
is maximum with an increment of about $85\%$ compared to T$0$ photoanode. The effect of porosity on the cells has also
improved the current density of the cells so that J$_{\rm SC}$ has increased from $12.59$ mA/cm$^2$ for T$0$ cell to $17.73$ mA/cm$^2$ for
T$3$ cell.
This improvement of J$_{\rm SC}$ stems from more dye loading and light scattering
due to the presence of the porosity in the photoanode.

%\begin
%\Table{\label{tabl1}Photovoltaics parameters of the DSSCs with different amounts of carbon nanospheres mixed with TiO$_2$.}
%\br
%&&&\centre{2}{Separation energies}\\
%\ns
%&T0&&\crule{2}\\
%Nucleus&(mg\,cm$^{-2}$)&Composition&$\gamma$, n (MeV)&$\gamma$, 2n (MeV)\\
%\mr
%$^{181}$Ta&$19.3\0\pm 0.1^{\rm a}$&Natural&7.6&14.2\\
%$^{208}$Pb&$\03.8\0\pm 0.8^{\rm b}$&99\%\ enriched&7.4&14.1\\
%$^{209}$Bi&$\02.86\pm 0.01^{\rm b}$&Natural&7.5&14.4\\
%\br
%\end{tabular}
%\item[] $^{\rm a}$ Self-supporting.
%\item[] $^{\rm b}$ Deposited over Al backing.
%\end{indented}
%\end{table}

\begin{table*}[h] 
\caption{\label{table}Photovoltaics parameters of the DSSCs with different amounts of carbon nanospheres mixed with TiO$_2$.}
\begin{indented}
    \item[]
\begin{tabular*}{0.7\textwidth}{@{\extracolsep{\fill} } l c  c  c  c  c  c  r |r |r |r |r |r | }
  \br
%  \hline
  Sample & T$0$  & T$1$ & T$2$ & T$3$ & T$4$ \\
  \hline
  J$_{\rm SC}$ (mA/cm$^2$) & $12.59$  & $14.99$  & $15.90$ & $17.73$ & $16.05$  \\
  
  V$_{\rm OC}$ (V) & $0.710$  & $0.685$  & $0.715$ & $0.720$ & $0.700$  \\
  
  FF & $0.64$  & $0.63$  & $0.61$ & $0.59$ & $0.607$  \\
  
  N$_{dye}$ ($10^6$ cm$^{\rm -2}$) & $4.44$ & $6.17$ & $6.91$ & $8.23$ & $6.11$  \\
  
  $\eta$ ($\%$) & $5.72$  & $6.47$  & $6.90$ & $7.59$ & $6.82$  \\
%  \hline
  \br 
    \end{tabular*}
  \end{indented}
\end{table*}

The porosity has a slight effect on the open circuit voltage (V$_{\rm OC}$) as it is seen in Table~\ref{table}. 
In the following, when we present the results of EIS analysis, we will describe these small changes in V$_{\rm OC}$ as a
consequence of charge recombination with electrolyte.

Now, we focus on the fill factor results presented in Table~\ref{table}.
The fill factor behaviour can be easily justified based on the change in current density. The increment of current density
causes high electron concentration leading to a greater resistance of the cells and therefore decreases the fill factor. 

\begin{figure}[b]
  \begin{center}
    \includegraphics[angle=-90,width=9cm]{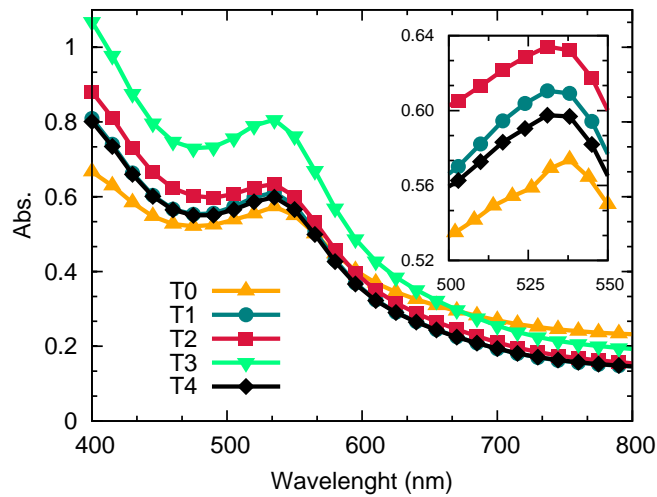}
    \caption{UV-Vis spectra of the dye-loaded TiO$_2$ layers T$0$, T$1$, T$2$, T$3$, and T$4$ photoanodes.}
    \label{uv}
  \end{center}
\end{figure}

To determine the absorbance of the dye-loaded porous TiO$_2$ layers, UV-Vis spectra were measured.
\fref{uv} shows UV-Vis spectra of the dye-sensitized T$0$, T$1$, T$2$, T$3$, and T$4$ photoanodes.
As seen in \fref{uv}, light absorbance by the dye-sensitized TiO$_2$ increases
from T$0$ up to T$3$ sample and decreases from T$3$ to T$4$.

\begin{figure}[t]
  \begin{center}
    \includegraphics[angle=-90,width=9cm]{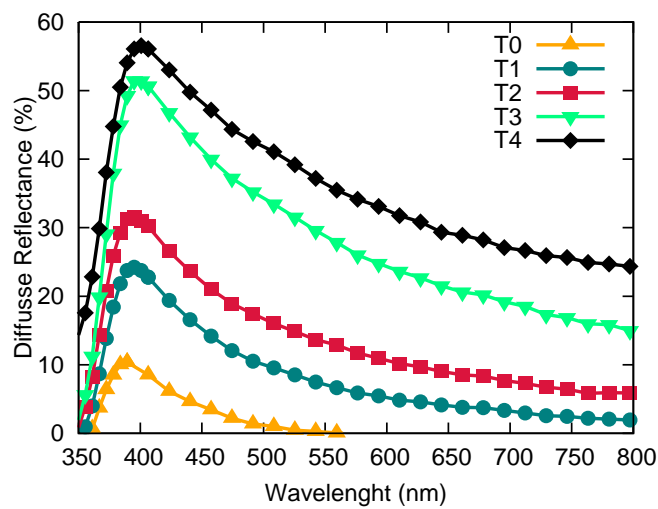}
    \caption{Diffuse reflectance spectra of T$0$, T$1$, T$2$, T$3$, and T$4$ photoanodes.}
    \label{Drs}
  \end{center}
\end{figure}

Diffuse reflectance was measured in order to investigate optical properties of T$0$, T$1$, T$2$, T$3$, and T$4$ electrodes
in the absence of dye sensitization.
Diffuse reflectance spectrum is an effective analysis for indicating the light scattering ability of samples.
\fref{Drs} clearly reveals the effect of the carbon nanospheres on diffuse
reflectance. As shown in \fref{Drs}, the ability of light scattering of the samples increases by increasing of the 
weight ratio of carbon nanospheres mixed with TiO$_2$. Increasing large cavities of size range $400$-$600$ nm
can be ideal for light scattering in visible region. 
\fref{Drs} indicates that the light scattering monotonically increases
from T$0$ to T$4$. On the other hand, 
the light absorbance by dye plotted in \fref{uv} becomes maximum for T$3$ 
and decreases by increasing the porosity of TiO$_2$ photoanode further.
This behaviour can be explained based on the size of carbon nanospheres. Because the carbon nanospheres have the size range
of $100$-$600$ nm, the number of large cavities increase by
increasing the weight ratio of carbon nanospheres.
Consequently, the light scattering continues to increase from T$3$ to T$4$ while dye loading starts to decrease beyond
the optimum amount of porosity $3$ $\%$ Wt..

\begin{figure}[b]
  \begin{center}
    \includegraphics[angle=0,width=7.5cm]{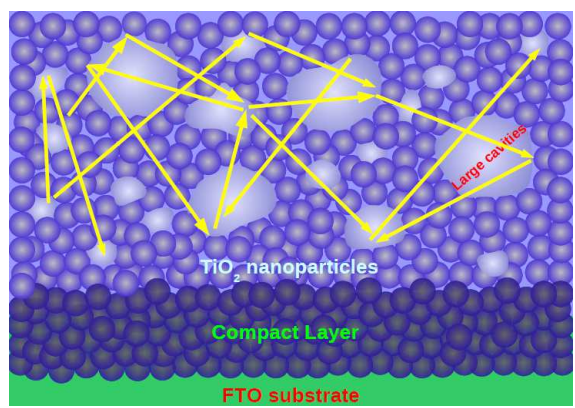}
    \caption{Schematic diagram of light scattering in TiO$_2$ photoanodes.}
    \label{Schem}
  \end{center}
\end{figure}

\fref{Schem} schematically shows how the presence of
large cavities leads to the light scattering and enhances the optical path
length in photoanode. These cavities increase multiple scattering of light and improve optical
properties of the cell (better light trapping and less transmittance). Small cavities make high surface area which lead to
more dye loading and light harvesting.

\begin{figure}[h]
  \begin{center}
    \includegraphics[angle=-90,width=9cm]{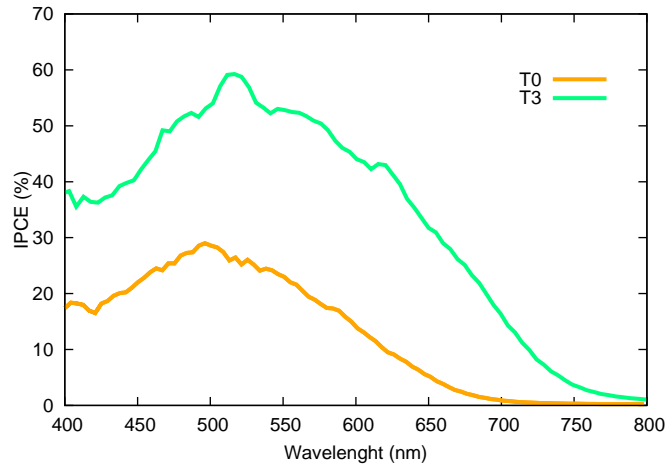}
    \caption{Incident photon to current conversion efficiency for T$0$ and T$3$ cells.}
    \label{IPCE}
  \end{center}
\end{figure}

The incident photon-to-current conversion efficiency (IPCE) is plotted in \fref{IPCE}
for T$0$ and T$3$ cells. This figure demonstrates that T$3$ cell has a higher IPCE than T$0$ cell
confirming a higher short circuit current density J$_{\rm SC}$ for T$3$ compared to T$0$ as it was also found
from I-V figure (\fref{Efficiency}).
It is also seen that the peak of IPCE curve is shifted from $30\%$ to $60\%$ and
the IPCE spectrum is broadened over the $400$-$600$ nm wavelength region which are due to the more dye loading and
light scattering in T$3$ photoanode.

\begin{figure}[t]
  \begin{center}
    \includegraphics[angle=-90,width=9cm]{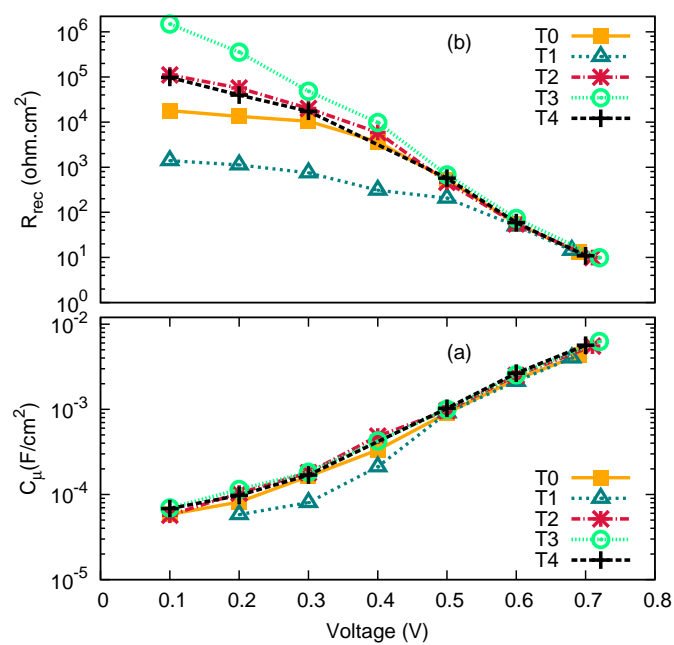}
    \caption{The chemical capacitance (a) and the recombination resistance (b) for T$0$, T$1$, T$2$, T$3$, and T$4$ cells
    obtained from EIS measurements in the dark conditions}
    \label{candr}
  \end{center}
\end{figure}
The EIS measurements were carried out to investigate the electron recombination and the band structure in fabricated cells.
The results of EIS measurements were fitted using previously developedmodel. From this fitting the chemical capacitance C$\mu$ and
recombination resistance R$_{\rm rec}$ were found. The results for C$\mu$
and R$_{\rm rec}$ are plotted versus voltage in \fref{candr}.

As seen in \fref{candr}a, the porosity in TiO$_2$ photoanode has no considerable effect on the chemical capacitance C$\mu$.
Since the slope of C$\mu$ reflects the TiO$_2$ density of states~\cite{Samadpour}, it can be concluded that the TiO$_2$ density
of state distributions are identical for all the different porosity.
Also there is no shift in the chemical capacitance of various cells and consequently no displacement for the TiO$_2$ conduction
band edge has occurred.

The recombination resistance can be used as a criterion for the recombination rate
so that a larger rate indicates a lower resistance and vice versa. \fref{candr}b shows that T$3$ and T$1$ cells have the
maximum and minimum recombination resistance, respectively. This behaviour can justify the maximum and
the minimum values of V$_{\rm OC}$ that we found from I-V analysis, see Table~\ref{table}.
Considering that increase of recombination reduces V$_{\rm OC}$, a maximum and a
minimum V$_{\rm OC}$ is plausible for T$3$ and T$1$ cells.
The seemingly disputed V$_{\rm OC}$ result that we found for T$0$ and T$1$ samples
can also be justified by taking into account the role of recombination. Although the T$1$ sample in contrast to the T$0$ sample
has a higher J$_{\rm SC}$ and a higher dye loading but the increase in the recombination
shown in \fref{candr}b leads to a lower V$_{\rm OC}$ for T$1$ compared to T$0$.

The recombination rate of T$2$ and T$4$ cells are almost the same as seen
from \fref{candr}b. This is compatible with the rather equal photovoltaic
parameters, see Table~\ref{table}, that we find from I-V figure for T$2$ and T$4$ cells.

\section{Conclusion}
The use of carbon nanospheres mixed with TiO$_2$ nanoparticles in DSSC photoanode with a facile method
demonstrated a higher photovoltaic performance compared to the TiO$_2$ without any carbon
nanospheres. By removing the carbon nanospheres, large and small cavities were created in the photoanode and enhanced the recombination resistance.
The large cavities improved the light scattering and therefore an increase in the optical path was achieved. On the other hand
the small cavities increase dye loading in the photoanode. 
These results are proven by DRS, desorption of dyes, and EIS
measurements. As a consequence an enhancement of about $40\%$ for current density and about $33\%$ for the cell efficiency were obtained.
Improving the photovoltaic performance with the method described in this article can be promising to
fabricate high efficiency dye-sensitized solar cells.

\ack
It is a pleasure to thank Sh. Dadgostar and F. Tajabadi for stimulating 
discussions. We also thank R. Mohammadpour and R. Ghahari for helpful
hints. E. Bayatloo is indebted to M. Samadpour for useful comments about
EIS analysis. We are grateful to R. Poursalehi for
reading the initial version of the manuscript and making useful suggestions.

\section*{References}
\bibliographystyle{iopart-num.bst}
%\bibliography{corr}
%\begin{thebibliography}
\providecommand{\newblock}{}

%\end{thebibliography}

\end{document}